\documentclass[onecolumn,prx,superscriptaddress]{revtex4-1}

\usepackage[dvips]{graphicx} 
\usepackage{amsmath,amssymb,amsthm,mathrsfs,amsfonts,dsfont}
\usepackage{enumerate}
\usepackage{epsfig}
\usepackage{subfigure}
\usepackage{xcolor}
\usepackage{amsthm}
\usepackage{framed}
\usepackage{braket}

\begin{document}
\title{Unified framework for quantumness --- coherence, discord, and entanglement}
\author{Xiao Yuan}
\affiliation{Center for Quantum Information, Institute for Interdisciplinary Information Sciences, Tsinghua University, Beijing, 100084 China}
\author{Hongyi Zhou}
\affiliation{Center for Quantum Information, Institute for Interdisciplinary Information Sciences, Tsinghua University, Beijing, 100084 China}
\author{Mile Gu}
\affiliation{School of Physical and Mathematical Sciences, Nanyang Technological University, Singapore 639673, Republic of Singapore}
\affiliation{Complexity Institute, Nanyang Technological University, Singapore 639673, Republic of Singapore}
\affiliation{Centre for Quantum Technologies, National University of Singapore,
3 Science Drive 2, Singapore, Republic of Singapore}

\author{Xiongfeng Ma}
\affiliation{Center for Quantum Information, Institute for Interdisciplinary Information Sciences, Tsinghua University, Beijing, 100084 China}

\begin{abstract}
From an operational perspective, quantumness characterizes the exotic behavior in a physical process which cannot be explained with Newtonian physics. There are several widely used measures of quantumness, including coherence, discord, and entanglement, each proven to be essential resources in particular situations. There exists evidence of fundamental connections amongst the three measures. However, those quantumnesses are still regarded differently and such connections are yet to be elucidated. Here, we introduce a general framework of defining a unified quantumness with an operational motivation founded on the capability of interferometry. The quantumness appears differently as coherence, discord, and entanglement in different scenarios with local measurement, weak reference frame free measurement, and strong reference frame free measurement, respectively. Our results also elaborate how these three measures are related and how they can be transformed from each other. This framework can be further extended to other scenarios and serves as a universal quantumness measure.
\end{abstract}

\maketitle

\section{Introduction}
Quantumness may come under in various shapes in different scenarios. One of the earliest forms of quantumness is the coherent superposition of a single quantum system. Quoting Einstein, \emph{God does not play dice with the world}, it is nowadays widely believed that the intrinsic randomness in a quantum measurement is a key feature that distinguishes the quantum theory from classical ones. From the perspective of quantum resource, intrinsic randomness comes from breaking the coherence of quantum states \cite{born1926quantentheorie}. The superposition or interference of distinguishable states -- coherence -- thus can be regarded as a mark for single partite quantumness \cite{Baumgratz14}.

In the past few decades or so, quantum information theory has been well developed. In quantum information processing, the existence of quantumness has been witnessed by specific tasks that can be fulfilled by quantum processes but not any classical process. For instance, Bell inequalities, satisfied by the classical theory, can be violated with certain quantum settings \cite{bell1964einstein,CHSH}; Quantum correlations enable extending secret keys between two remotely separated users \cite{bb84, Ekert91}, which is impossible with classical processes; Quantum computing can tackle classically intractable problems \cite{Shor97}. In these and many other tasks, entanglement, which measures a special form of correlation of multipartite quantum systems, has been recognized as the central element that is responsible for the advantage of quantum process \cite{Horodecki09}. Entanglement becomes the most widely used measure of quantumness \cite{Bennett96, Vedral98}.

Beside entanglement, another important quantumness in quantum information processing is discord. As a general measure for multipartite quantum correlation, discord plays important roles in many tasks including quantum computing \cite{Datta09}, remote state preparation \cite{dakic2012quantum}, quantum metrology \cite{Girolami13}, and others \cite{gu2012observing}. In those tasks, the quantum advantage can be explained by discord even in the absence of entanglement.

The three quantumness measures, coherence, discord, and entanglement, are the most widely used ones, for which, there are resource frameworks \cite{streltsov2016quantum, Horodecki09, Modi12} that describe how they can be characterized, manipulated, and quantified, respectively. There are many examples indicating fundamental connections between the three measures. For quantum correlation, including entanglement and discord, it is shown that all nonclassical correlations can be activated into distillable entanglement \cite{Piani11, Streltsov11}. Recently, enormous efforts have been devoted to investigate the relation between coherence and quantum correlation. For instance, the trade-off between coherence and correlation measures has been analyzed under different scenarios \cite{Bera15,Svozil15,Yao15,Bagan16,Radhakrishnan16, Streltsov17}. Also, considering incoherent operations, it is shown that coherence can be converted into quantum correlation \cite{Streltsov15, Ma16}. In addition, coherence and quantum correlation are shown to play important roles in several information tasks, such as frozen quantumness \cite{Maziero09,Bromley15,yu2016measure} and quantum state merging \cite{horodecki2005partial,Horodecki2007,Streltsov16}.

From these observations, we see that coherence, discord, and entanglement are deeply connected concepts.
Our work supports this intuition by proposing a unified framework of quantumness by considering a simple information task.
A challenge of this unification lies on the fact that coherence characterizes quantumness of a single system while entanglement and discord characterize multipartite quantum correlations. In addition, difficulties also stem from that coherence is defined on a specific measurement basis while entanglement and discord is independent of (local) basis. Furthermore, although discord and entanglement both describe multipartite correlation, their similarity and difference are not fully understood.

Tracing back the origin of the mystery of quantum mechanics, it is the wave-particle duality that first confuses physicists, including Einstein. From today's viewpoint, Einstein's quote indicates that he disagreed with Born's probability interpretation of wave functions \cite{born1926quantentheorie}. It has been long well-known, before the birth of quantum mechanics, that the wave property of a physical subject can be demonstrated by interference. In fact, it is Young's double-slit experiment confirms the wave property of light. We follow this track to unify various notions of quantumness. In our framework, we operationally identify quantumness based on the capability of interferometry. Under different scenarios, we show that the associated quantumness is coherence, discord, and entanglement.

\subsection{Preliminaries}
We review the definition of a general quantumness framework, which usually relies on identifying classical states and classical operations. Focusing on an operational task, the corresponding quantumness is witnessed when quantum behavior that cannot be explained classically is observed. A state $\sigma$ is called \emph{classical} when it exhibits no quantum behavior. Denote the set of classical states by $\mathcal{C} = \{\sigma\}$, then a state $\rho$ that does not belong to $\mathcal{C}$ is called \emph{quantum}. Based on classical states, an operation $\Phi^\mathcal{C}$ is classical when it is physically realizable and cannot generate quantum state from any classical state. We leave a rigourous definition of classical operations in Methods.
With classical states and operations, a resource framework of quantumness is completed by defining measures, which is a real-valued function of states, $Q(\rho)$. Generally, a quantumness measure should satisfy the monotonicity requirement that classical operations cannot increase quantumness. Note that, we focus on the quantumness of states. A general quantumness framework for processes can be similarly defined by utilizing the channel-state duality \cite{CHOI1975285}.

In this work, we concentrate on four different quantumness, coherence \cite{Baumgratz14}, basis-dependent (BD) discord \cite{Ollivier01}, discord \cite{Henderson01,Ollivier01}, and entanglement \cite{Bennett96}, of which BD-discord plays as a bridge that links coherence and quantum correlation. Focusing on different scenarios, distinct quantumness can be defined based on different set of classical states and operations. In Supplementary Materials, we summarize the existing frameworks for these quantumness. In the following, we will show our unified framework, in which these quantumness naturally arises.



\section{Theory}
\subsection{Double slit experiment}
\begin{figure}[hbt]
\centering \resizebox{8cm}{!}{\includegraphics{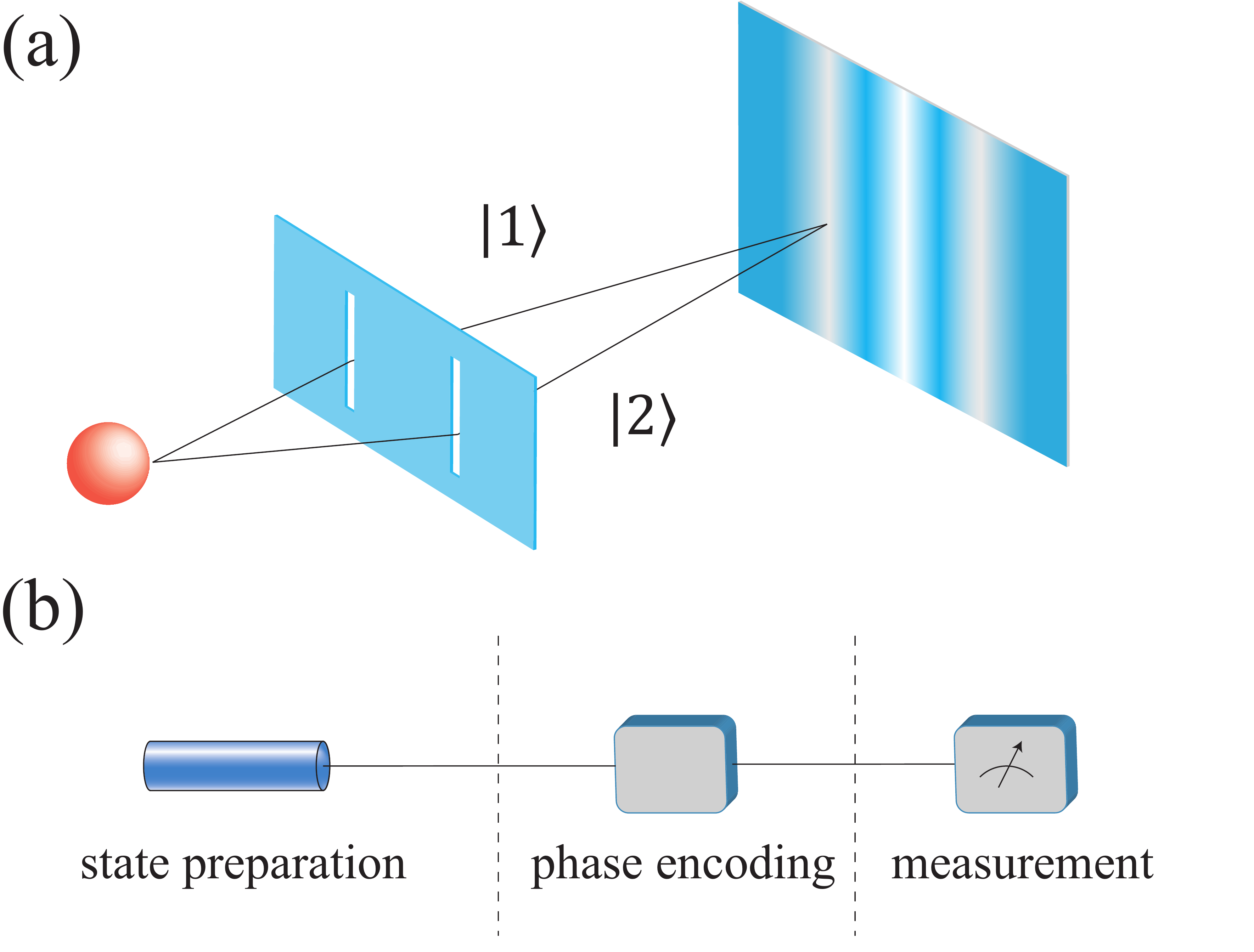}}
\caption{Double slit experiment of an election. (a) Interference pattern is observed when the electron is in a superposition of the path basis $J = \{\ket{1}, \ket{2}\}$. (b) A general interferometric process that consists of state preparation, phase encoding, and measurement. The quantumness of the prepared state is witnessed when interference pattern, i.e., nonzero phase information, is observed from the measurement outcome.} \label{Fig:interference}
\end{figure}

As an illustrative example, we consider the double slit experiment of an electron, shown in Fig.~\ref{Fig:interference}(a), as our first operational task. Classically, the electron will go through either path $\ket{1}$ or $\ket{2}$ and display no interference pattern. While, when the electron is in a superposition of two paths, the quantum behavior of interference can be observed. Schematically, see Fig.~\ref{Fig:interference}(b), the double slit experiment can be regarded as an interferometric process that probes the phase difference between different paths. Considering this superposition as a quantum feature, while a mixture of two paths as classical states, the interferometric capability is thus a traditional signature of the quantumness of superposition.

\subsection{An interferometric framework}
The double slit experiment can be generalized to observe phase behavior on $d_A$ paths (labeled by $A$) and meantime the particle potentially has $d_B$ internal degrees of freedom (labeled by $B$). In this case, we focus on a $d_A\times d_B$-dimensional Hilbert space $\mathcal{H}_{AB} = \mathcal{H}_{A}\otimes\mathcal{H}_{B}$ with bases $J_A=\{\ket{j_A}\}_{j_A=1,2,\dots,d_A}$ and $J_B=\{\ket{j_B}\}_{j_B=1,2,\dots,d_B}$ for path and internal degrees of freedom, respectively. In principle, system $A$ and $B$ can also be regarded as two individual subsystems.

An interferometric process generally consists of three stages, state preparation, phase encoding, and measurement. After state preparation, an initial state $\rho_{AB}$ from $\mathcal{H}_{AB}$ is prepared. Denote $\phi_{j_A}$ as the accumulated phase for path $\ket{j_A}$, the phase encoding process can be described by a unitary operation,
\begin{equation}\label{Eq:encoding}
U_{J_A,\phi_A} = \sum_{j_A = 1}^{d_A} e^{-i\phi_{j_A}}\ket{j_A}\bra{j_A}\otimes I_B,
\end{equation}	
with ${\phi}_A = (\phi_1, \phi_2, \dots,\phi_{d_A})$ and $I_B$ being the identity matrix of the internal degrees of freedom.
After phase encoding, the state evolves to $\rho_{AB}' = U_{J_A,\phi_A}\rho_{AB} U_{J_A,\phi_A}^\dag$. In the measurement phase, we consider a general positive-operator valued measure (POVM) $\{M_1, M_2, \dots, M_D|\sum_iM_j = I, M_j\ge0\}$ on $\rho_{AB}'$, where $D$ is the number of POVM elements. The measurement outcome is denoted as a random variable $X_M$.

Under the generalized interferometric process, the non-classical or quantum behavior is defined by the \emph{interferometric capability}, i.e., the ability of probing the phase information $\vec{\phi}_A$. A state $\rho_{AB}$ is called classical when the measurement outcome $X_M$ is independent of the phase information $\vec{\phi}_A$, i.e., $I(X_M, \vec{\phi}_A) = 0$, for any possible measurement. Here, we consider $\vec{\phi}_A$ as a random variable and $I(X,Y)$ is the mutual information of two random variables. On the other hand, a state is considered non-classical or quantum if one can acquire nonzero information of the phase with a proper measurement.

\subsection{Adversarial scenario}
The interferometric capability generally relies on the encoding basis or the reference frame, which in practice can vary with time or be difficult to acquire under a black-box phase encoding scenario \cite{Girolami13}. Such a practical issue is equivalent to the worst case scenario where an adversary controls the phase encoding basis according to her local information $\rho_E$ to minimize the phase information that can be learned from the measurement result. It is thus also interesting to investigate the interferometric capability, i.e., quantumness, of quantum states under adversary's control. In such an adversarial scenario, a state $\rho_{AB}$ is called quantum only when the measurement outcome $X_M$ has nonzero phase information, i.e., $\min_E I(X_M, \vec{\phi}_A) > 0$, where the minimization is over all possible manipulations by the adversary, who may share entanglement with $\rho_{AB}$ and control the phase encoding basis as described below.

\begin{figure}[hbt]
\centering \resizebox{8cm}{!}{\includegraphics{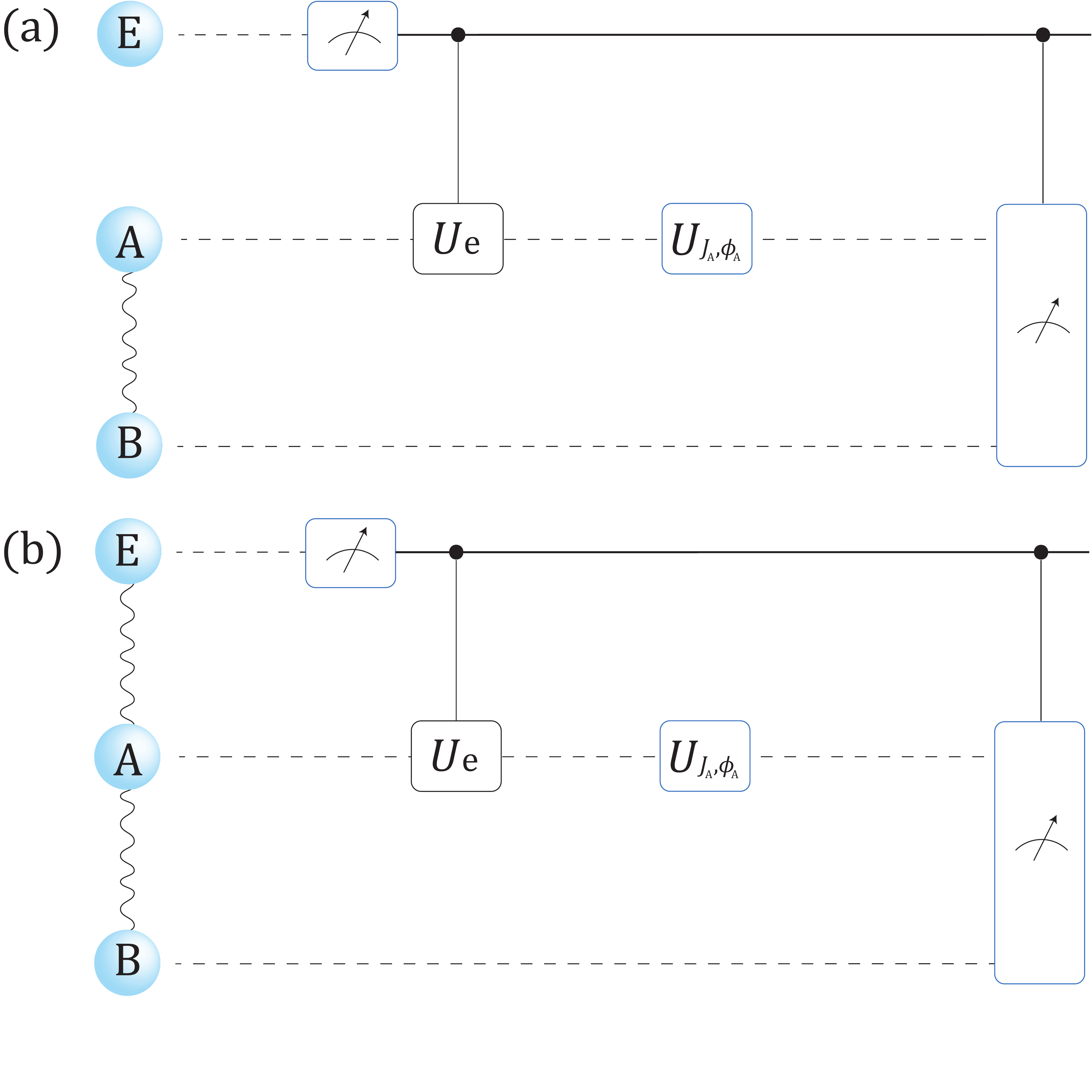}}
\caption{Interferometric process under adversarial scenarios. (a) The weak adversary is not entangled with system $AB$ while (b) the strong adversary shares maximal entanglement with system $AB$. The adversary measures her local system to generate a basis choice $e$, based on which she rotates the phase encoding basis $J_A$ to an arbitrary one $J_A^e=\{\ket{j_A^e} = U_e^\dag\ket{j_A}\}$ by acting a unitary operation $U_e = \sum_{j_A}\ket{j_A}\bra{j_A^e}$ on system $A$. That is, $\rho^e_{AB} = U_{J_A,\phi_A}U_e\rho_{AB}U_e^\dag U_{J_A,\phi_A}^\dag = U_eU_{J_A^e,\phi_A}\rho_{AB} U_{J_A^e,\phi_A}^\dag U_e^\dag$ with $U_{J_A^e,\phi_A}^\dag = U_e^\dag U_{J_A,\phi_A}U_e$.
Note that, strong adversary is strictly more powerful than weak adversary. } \label{Fig:entangled}
\end{figure}

We consider that the adversary first measures her local system $\rho_E$ to generate a basis choice $e$; then she rotates the phase encoding basis to $J_A^e=\{\ket{j_A^e} = U_e^\dag\ket{j_A}\}$ by applying the rotation $U_e^\dag$ based on $e$. We assume that the measurement outcome $e$ is revealed to the interferometry measurement. Otherwise, it is not hard to see that the adversary can always destroy the interferometry capability for any input state $\rho_{AB}$.
In this work, we consider two different ways that the adversary generate the basis choices, as shown in Fig.~\ref{Fig:entangled}.
The adversary is called  \emph{weak} when her local system $E$ is not entangled with system $AB$ and \emph{strong} when system $ABE$ is maximally entangled.
As the phase encoding reference frame is unknown for each basis choice, we identify such interferometric capability by weak and strong reference frame free (RFF) quantumness for weak and strong adversaries, respectively.

Before presenting our result, we first briefly summarize the definitions of the four quantumness measures.

\subsection{Four quantumness measures}
Coherence is defined for a single quantum system on a specific measurement basis \cite{Baumgratz14}.
Considering the space of path and the phase encoding basis $J_A$, a state $\sigma^{J_A}_A$ is called incoherent state when
\begin{equation}\label{Eq:sigmaA}
  \sigma^{J_A}_A = \sum_{j_A=1}^{d_A} p_{j_A}\ket{j_A}\bra{j_A},
\end{equation}
and coherent state otherwise.

For BD-discord  \cite{Ollivier01}, we consider the joint state $\rho_{AB}$ of path and internal degrees of freedom. A state has no BD-discord on $J_A$ iff
\begin{equation}\label{Eq:sigmaAB}
  \sigma^{J_A}_{AB} = \sum_{j_A=1}^{d_A} p_{j_A}\ket{j_A}\bra{j_A}\otimes \rho^{j_A}_B,
\end{equation}
where $p_{j_A}\ge 0$, $\sum_{j_A} p_{j_A}=1$, and $\rho^{j_A}_B$ is an arbitrary state from $\mathcal{H}_B$.

Quantum correlation also defines the quantumness of joint system. The set of states that have zero discord \cite{Henderson01,Ollivier01} is defined by the union of zero BD-discord state of all local bases,
\begin{equation}\label{Eq:cqstate}
  \mathcal{C} = \bigcup_{J_{A}} \{\sigma^{J_A}_{AB}\},
\end{equation}
where $\sigma^{J_A}_{AB}$ is defined in Eq.~\eqref{Eq:sigmaAB} for basis $J_A$.

A state that has no entanglement is called separable state, which is given by
\begin{equation}\label{eq:separable}
 \sigma_{AB} = \sum_{j}^{} p_{j}\rho^{j}_A \otimes \rho^{j}_B.
\end{equation}


\subsection{Our result---classical states}
Now, we show that coherence, BD-discord, discord, and entanglement are necessary and sufficient resources for demonstrating the interferometric capability in different scenarios. Here we only discuss the necessary argument and leave the rigouros derivations in Supplementary Materials.

First, we consider the interferometric process without the presence of adversary. When the Hilbert spaces of path and internal degrees of freedom are uncorrelated, the input state can be expressed as $\rho_{AB} = \rho_A\otimes\rho_B$ and the interferometric ability is independent of the internal degrees of freedom. Focusing on the path state,

\emph{Result 1.}---State $\sigma_A$ displays no interferometric capability (i.e., $I(X_M, \vec{\phi}_A) = 0, \forall M$) iff $\sigma_A$ is incoherent state on basis $J_A$, see Eq.~\eqref{Eq:sigmaA}.

The intuition is that when the input state $\sigma_A$ is incoherent on the phase encoding $J_A$, the state after phase encoding, i.e., $\sigma_{A}' = U_{J_A,\phi_A}\sigma_{A} U_{J_A,\phi_A}^\dag$ is identical to the input state $\sigma_A$ hence independent of the phase information. On the other hand, as long as the input state has nonzero coherence, phase information can be encoded and read out by a proper measurement.
Considering a specific example where only a finite number of different phases are chosen, the probability of guessing the phase information correctly is quantitatively characterized by the robustness of coherence. We refer to Ref.~\cite{Napoli16} for details.
In general, when the input state fields correlation between the path and internal degrees of freedom, we can similarly prove that

\emph{Result 2.}---State $\sigma_{AB}^{J_A}$  displays no interferometric capability iff $\sigma_{AB}^{J_A}$ has zero BD-discord on basis $J_A$ as defined in Eq.~\eqref{Eq:sigmaAB}.


In the presence of adversary, correlation between the path and internal degrees of freedom is necessary for displaying quantum behavior.
Under a weak adversary, we can effectively consider that there is one but unknown phase encoding basis $J_A'=\{\ket{j_A'}\}$. In this case, an uncorrelated state $\sigma_{AB} = \rho_{A}\otimes\rho_B$ becomes classical when $J_A' = \{\ket{j_A'}\}$ is chosen in which $\rho_A$ has a spectral decomposition $\rho_A = \sum_{j_A'} \lambda_{j_A'} \ket{j_A'}\bra{j_A'}$. In general, states with zero discord, as defined in Eq.~\eqref{Eq:cqstate}, displays zero interferometric capability under a weak adversary. This is because the adversary can always choose a phase encoding basis $J_A'$ in which a zero discord state also has zero BD-discord. Considering a strong adversary, who holds a purification of $\ket{\phi}_{ABE}$ with $\rho_{AB}=\mathrm{tr}[\ket{\phi}_{ABE}\bra{\phi}_{ABE}]$. She can rotate the phase encoding basis according to the measurement result on her local quantum system $E$. Suppose the adversary performs a local measurement on $E$, which effectively collapses the remaining system to a decomposition of $\rho_{AB}=\sum_ep_e\ket{\psi_{AB}}_e\bra{\psi_{AB}}_e$, she can thus rotate the measurement basis to $J_A^e$ individually for each measurement outcome $e$. Therefore, as long as $\rho_{AB}$ can be decomposed into a convex combination of BD-discord states of all measurement bases, i.e. a separable state defined in Eq.~\eqref{eq:separable}, it cannot be used for interferometry under a strong adversary.

\emph{Result 3 (discord) \& 4 (entanglement).}---State $\sigma_{AB}$ displays no interferometric capability under weak and strong adversaries iff $\sigma_{AB}$ has zero discord (Eq.~\eqref{Eq:cqstate}) and entanglement (Eq.~\eqref{eq:separable}), respectively.



\subsection{Quantumness measures}
\begin{figure}[bht]
\centering
\resizebox{8cm}{!}{\includegraphics[scale=1]{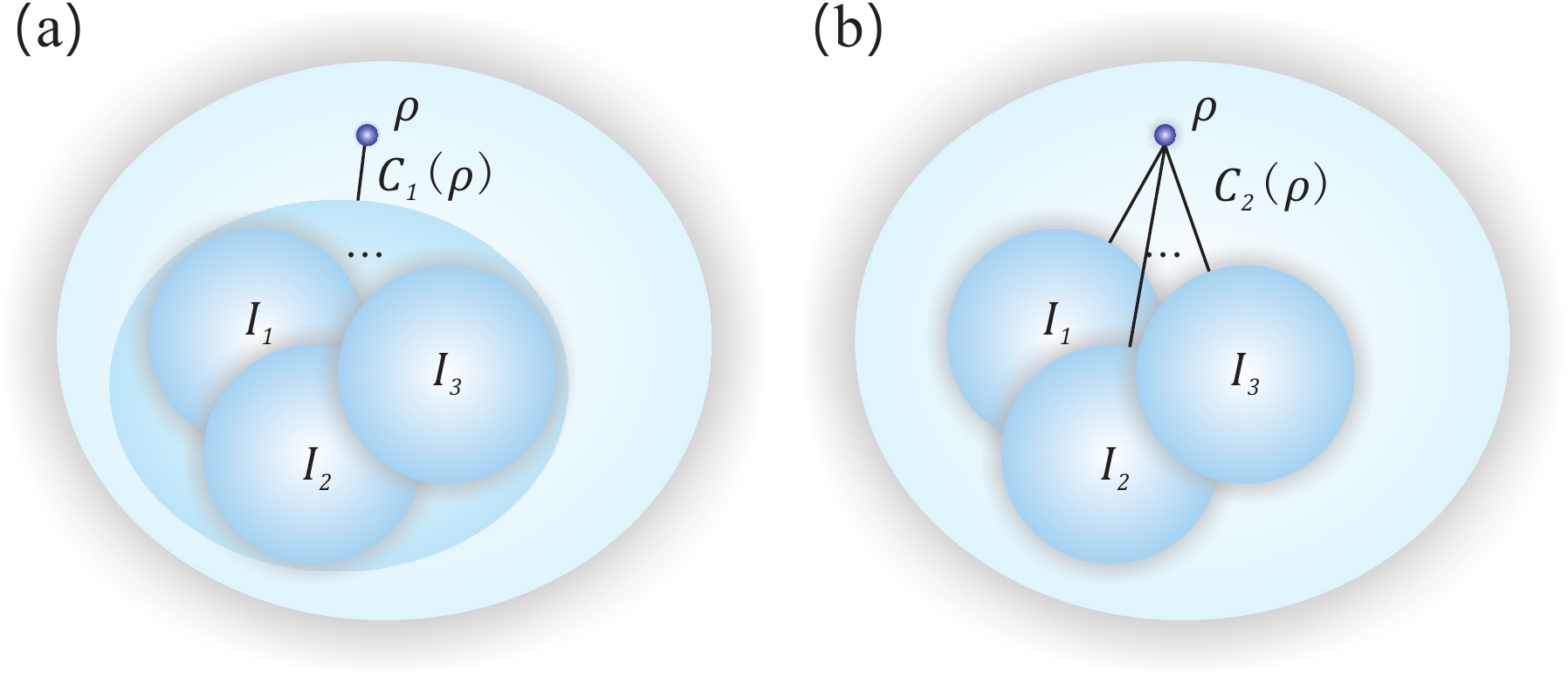}}
\caption{Two ways of defining RFF quantumness. (a) Classical states $\sigma_{AB}$ are defined by convex combinations of zero BD-discord states $\sigma^{J_A}_{AB}$ all all basis $J_A$, $\sigma_{AB}=\sum_ep_e\sigma^{J_A}_{AB}$. (b) The set of classical states $\mathcal{C}$ are the union of all zero BD-discord states, $\mathcal{C} = \bigcup_{J_{A}} \{\sigma^{J_A}_{AB}\}$.}\label{fig:Twoways}
\end{figure}

Based on the definitions of classical states and classical operations, quantumness measures naturally arise. As coherence and BD-discord are defined on a local basis of single partite and bipartite states, respectively, the quantumness measures can be similarly defined as proven in Supplementary Materials. Furthermore, classical states of discord and entanglement are illustrated in Fig.~\ref{fig:Twoways}. That is, zero-discord (entangled) states are the union (convex combinations) of zero BD-discord states of all local bases. 

Considering the minimal distance between the target and all classical states as a measure of quantumness, we can define a discord measure by minimizing over all local basis of a BD-discord measure.
Denote the BD-discord measure of $\rho_{AB}$ on local basis $J_A$ to be $Q_{J_A}(\rho_{AB})$, a discord measure can be defined by
\begin{equation}\label{Eq:RFFQ}
  Q_D(\rho_{AB}) = \min_{J_A} Q_{J_A}(\rho_{AB}).
\end{equation}
Interestingly, even though the BD-discord measure $Q_{J_A}(\rho_{AB})$ is not a distance measure, we prove it to be a discord measure in Supplementary Materials. For a strong adversary, we consider a purification state $\ket{\phi}_{ABE}$ with $\rho_{AB}=\mathrm{tr}[\ket{\phi}_{ABE}\bra{\phi}_{ABE}]$. Based on a measurement of $E$, which essentially decides a decomposition of $\rho_{AB}$, Eve can rotate the phase encoding basis to minimize the quantumness. Hence,

\emph{Result 6.}---The quantumness measure against strong adversary
\begin{equation}\label{Eq:RFFEntanglement}
  Q_E(\rho_{AB}) = \min_{p_e,\ket{\psi_{AB}}_e}\sum_e p_e\min_{J_A} Q_{J_A}(\ket{\psi_{AB}}_e)
\end{equation}
is an entanglement measure of $\rho_{AB}$.

\section{Examples: photonic setup}
\subsection{Ideal test}
\begin{figure*}[hbt]
\centering \resizebox{14cm}{!}{\includegraphics{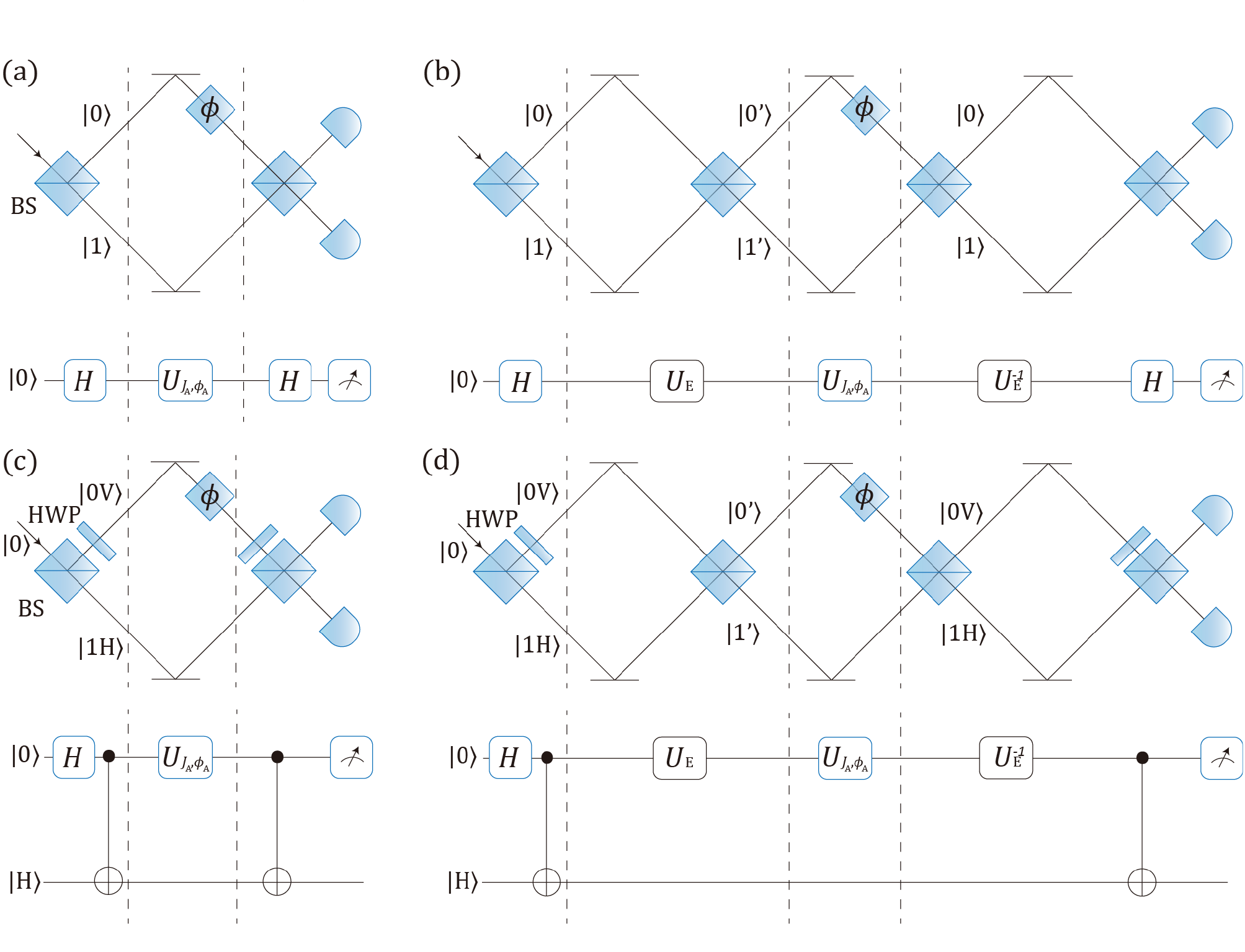}}
\caption{Photonic setups of four interferometric processes (a) assisted with coherence of paths; (b) under adversary's attacks; (c) with correlation between path and polarization; (d) with correlation and under adversary's attack. BS: beam splitter; $\ket{0}$, $\ket{1}$: two paths; HWP: half wave plate; $\ket{H}$: horizontal polarization; $\ket{V}$: vertical polarization; H: Hadamard gate;} \label{Fig:inter}
\end{figure*}

Here, we present a photonic setup for demonstrating the relation between quantumness and interferometric capability. Focusing on Fig.~\ref{Fig:inter}(a), we can test that the coherence on two paths $\{\ket{0}, \ket{1}\}$ is necessary for probing the phase $\phi$. When the beam splitter is replaced by a random switch, which selects the path according to a random bit, the prepared state will be in a mixture of the two paths and hence display no quantum effects. In Fig.~\ref{Fig:inter}(b), we consider that the phase encoding basis can be controlled by a weak adversary. Under this scenario, the phase information cannot be obtained when the adversary selects an appropriate basis. In Fig.~\ref{Fig:inter}(c), we consider interferometry with internal degrees of freedom, polarization. Even the local state of path contains no coherence, the correlation between path and polarization can still be used for probing the phase information. In Fig.~\ref{Fig:inter}(d), we consider in an adversarial scenario. In this picture, we can see that the quantum correlation of the prepared state guarantees the interferometric capability. 


In the experimental setup in Fig.~\ref{Fig:inter}, we only show the interferometric capability with ideal input states. In experiment, the phase information can also be obtained even with imperfect inputs. Next, we put such an argument into quantitative expression by showing a direct relation between phase information and the quantumness of the prepared state. Therefore, our result can also be experimentally tested in a quantitative way with practical input states.

\subsection{Quantitative comparison}
Generally, the phase information in the interferometric process can be quantified with Fisher information \cite{Wigner63}.
For simplicity, suppose there are only two paths $J_A=\{\ket{0},\ket{1}\}$ and the phase encoded on $\ket{0}$ and $\ket{1}$ are $0$ and $\phi_{A}$. Then the information encoding operation on system $A$ is $U_{J_A,\phi_{A}} = e^{-i\phi_A H_{J_A}}$, where $H_{J_A} = \ket{1}\bra{1}_{A}$. The skew information $\mathcal{I}(\rho_A, H_{J_A}) = -\mathrm{Tr}\{[\sqrt{\rho_A},H_{J_A}]^2\}/2$ that lower bounds \cite{Luo03} the fisher information of the phase information $\phi_A$ also measures the coherence \cite{Girolami2014Observable} in $\rho$. That is, a nonzero skew information $I(\rho_A, H_{J_A})$ indicates nonzero information of $\phi_A$ and hence coherence of $\rho_A$ on the $J_A$ basis. 
When considering joint measurement on system $AB$, skew information of local observable $H_A$ can be similarly defined by $\mathcal{I}_{\mathrm{BD}}(\rho_{AB}, H_{J_A}\otimes I_B) = -\mathrm{Tr}\{[\sqrt{\rho_{AB}},H_{J_A}\otimes I_B]^2\}/2$, which measures the BD-discord of $\rho_{AB}$. Furthermore, in the weak adversarial scenario, the weak RFF skew information can be defined by minimizing over all information encoding basis $\mathcal{I}_{\mathrm{D}}(\rho_{AB}) = \min_{J_A}\mathcal{I}_{\mathrm{BD}}(\rho_{AB}, H_{J_A}\otimes I_B)$. In literature, such RFF skew information has been shown a discord measure of quantum correlation \cite{Girolami13}. In the strong adversarial scenario, the basis can be controlled based on quantum information. Following the definition of Eq.~\eqref{Eq:RFFEntanglement}, we define the strong RFF skew information
\begin{equation}\label{}
  \mathcal{I}_{\mathrm{E}}(\rho_{AB}) = \min_{p_e,\ket{\psi_{AB}}_e}\sum_e p_e \min_{J_A}\mathcal{I}_{\mathrm{BD}}\left(\ket{\psi_{AB}}_e, H_{J_A}\otimes I_B\right),
\end{equation}
which measures the amount of information can be obtained under attacks of strong adversary.

\section{Discussion}
As summarized in Table~\ref{Table:measure}, we derive a unified framework for coherence, BD-discord, discord, and entanglement and show them as the resource for demonstrating the interferometric capability in different scenarios.
Furthermore, we present a photonic experimental setup to test our result. By quantitatively measuring the phase information, we show that the interferometric capability is measured by the quantumness of the input state. With imperfect input states, such direct relations can also be experimentally verified.
\begin{table*}[bht]\centering
\caption{Different measures for quantumness based on interferometry.}
\begin{tabular}{ccccccc}
  \hline
  {Quantumness} & System & Basis & {Adversary} & {Classical states} & Example\\
  \hline
  Coherence & $A$ & $J_A$ basis &No& Eq.~\eqref{Eq:sigmaA} & Fig.~\ref{Fig:inter}~(a)\\
  BD-discord & $AB$ & $J_A$ basis &No& Eq.~\eqref{Eq:sigmaAB} & Fig.~\ref{Fig:inter}~(c)\\
  Discord & $AB$ &RFF&Weak& Eq.~\eqref{Eq:cqstate} &Fig.~\ref{Fig:inter}~(d)\\
  Entanglement & $AB$ &RFF&Strong& Eq.~\eqref{eq:separable} & Fig.~\ref{Fig:inter}~(d)\\
  \hline
\end{tabular}\label{Table:measure}
\end{table*}

The unified framework of quantumness defined in this work is for discrete variable systems. In future works, we expect that it can be extended to the most general form of the quantumness of states, for instance, quantumness of continuous variable states. As we only focus on the quantumness of quantum states, while quantum measurement also plays important roles, we expect a similar definition for the quantumness of measurement can be proposed. Finally, our unification would shed light on a universal quantumness framework.

\section{Methods}
\subsection{Classical operation and quantumness measure}
The definition here follows from previous results of coherence \cite{Baumgratz14}, basis-dependent discord \cite{Ollivier01}, discord \cite{Henderson01,Ollivier01}, and entanglement \cite{Horodecki09}.
First, a classical operation $\Phi^C$ should be physically realizable, i.e., it is a completely positive trace preserving (CPTP) map. In addition, a classical operation cannot generate quantumness from classical states, i.e., $\Phi^C(\sigma)\in \mathcal{C}, \forall \sigma\in \mathcal{C}$. In Kraus representation, classical operation is defined by $\Phi^C(\sigma) = \sum_n\hat{K}_n\sigma\hat{K}_n^\dag \subset \mathcal{C}, \forall \sigma\in \mathcal{C}$, where $\{\hat{K}_n\}$ is a series of Kraus operators satisfying $\sum_n \hat{K}_n^\dag\hat{K}_n=I$.
Extra constraints can be added to the definition of classical operations. For instance, we can further require that classical operation cannot generate quantumness even under post-selection, $\hat{K}_n\sigma\hat{K}_n^\dag \subset \mathcal{C},  \forall \sigma\in \mathcal{C}$. 

A quantumness measure is a real-valued function of states, $Q(\rho)$, that satisfies the properties in Table.~\ref{Fig:measureProperties}. Extra requirement such as convexity can be added.
\begin{table}[htb]
\begin{framed}
\centering
\begin{enumerate}[(C1)]
\item
Vanishes for classical state: $Q(\sigma)=0,\forall \sigma\in \mathcal{C}$. Stronger condition: (C1') $Q(\sigma)=0$ iff $\sigma\in \mathcal{C}$;
\item
\emph{Monotonicity}: classical operation cannot increase quantumness, (C2a) $Q(\rho)\geq Q\left(\Phi^C(\rho)\right)$, (C2b) $Q(\rho) \geq \sum_n p_n C(\rho_n)$;
\end{enumerate}
\end{framed}
\caption{Properties of a generalized quantumness measure. } \label{Fig:measureProperties}
\end{table}

\emph{Acknowledgement.}---We acknowledge H.-K.~Lo and J.~Ma for the insightful discussions. This work was supported by the National Natural Science Foundation of China Grants No.~11674193.
\bibliographystyle{apsrev4-1}
\bibliography{QFFQbib}

\end{document}